\documentclass{aa}
\usepackage{graphics}
\usepackage{txfonts}
\usepackage{natbib}

\newcommand{\kms}   {km~s$^{-1}$}

\def\gs{\mathrel{\raise0.35ex\hbox{$\scriptstyle >$}\kern-0.6em
\lower0.40ex\hbox{{$\scriptstyle \sim$}}}}
\def\ls{\mathrel{\raise0.35ex\hbox{$\scriptstyle <$}\kern-0.6em
\lower0.40ex\hbox{{$\scriptstyle \sim$}}}}

\begin{document}

   \title{ A new probe of magnetic fields during high-mass star formation}
   \titlerunning{Zeeman splitting of 6.7~GHz methanol masers}

 \subtitle{Zeeman splitting of 6.7~GHz methanol masers}

   \author{W.H.T. Vlemmings\inst{1}
          }

   \offprints{WV (wouter@astro.uni-bonn.de)}

   \institute{Argelander Institute for Astronomy, University of Bonn,
     Auf dem H{\"u}gel 71, 53121 Bonn, Germany}

   \date{Received ; accepted }

% an abstract (must be included!):

   \abstract{The role of magnetic fields during high-mass star
     formation is a matter of fierce debate, yet only a few direct
     probes of magnetic field strengths are available.}{The
     magnetic field is detected in a number of massive star-forming regions
     through polarization observations of 6.7~GHz methanol
     masers. Although these masers are the most abundant of the maser
     species occurring during high-mass star formation, most magnetic
     field measurements in the high-density gas currently come from OH
     and H$_2$O maser observations.}{The 100-m Effelsberg telescope
     was used to measure the Zeeman splitting of 6.7~GHz methanol
     masers for the first time. The observations were performed on a
     sample of 24 bright northern maser sources.}{Significant Zeeman
     splitting is detected in 17 of the sources with an average
     magnitude of $0.56$~m~s$^{-1}$. Using the current best estimate
     of the 6.7~GHz methanol maser Zeeman splitting coefficient and a
     geometrical correction, this corresponds to an absolute magnetic
     field strength of $23$~mG in the methanol maser region.}{The
     magnetic field is dynamically important in the dense maser
     regions. No clear relation is found with the available OH maser
     magnetic field measurements. The general sense of direction of
     the magnetic field is consistent with other Galactic magnetic
     field measurements, although a few of the masers display a change
     of direction between different maser features. Due to the
     abundance of methanol masers, measuring their Zeeman splitting
     provides the opportunity to construct a comprehensive sample of
     magnetic fields in high-mass star-forming
     regions. }

   \maketitle

\section{Introduction}

Although massive stars play an important role in the chemical and
energetic evolution of their host galaxies, their formation mechanism
remains elusive. This problem is the topic of extensive observational
and theoretical efforts.  Even though few of the current simulations
include magnetic fields, the influence of magnetism on the star
formation processes is extensive as it can support a molecular cloud
against collapse, affect core fragmentation and change the feedback
processes \citep[e.g.][and references therein]{krumholz07}.

Most current high-mass star formation magnetic field information comes
from H$_2$O and OH maser polarization observations. The observations
of the H$_2$O maser Zeeman effect using Very Long Baseline
Interferometry (VLBI) reveal field strengths between $10$ and
$600$~mG, while the linear polarization measurements reveal a complex
but often ordered magnetic field morphology \citep[e.g.][]{Vlemmings06}.
Aside from H$_2$O masers, tracing high-density regions ($n_{\rm
  H_2}\approx10^8-10^{11}$~cm$^{-3}$), the magnetic field in the less
dense surrounding regions is typically probed by polarimetric OH maser
observations \citep[e.g.][]{Bartkiewicz05}. These observations reveal
fields of a few mG as well as ordered structure in the magnetic
field. However, the strongest and most abundant of the high-mass star
formation region masers arises from the 6.7~GHz $5_1 - 6_0A^+$
methanol transition, and for this maser hitherto only very few
polarization observations exist.

Over $95\%$ of the class II 6.7~GHz masers have been found to harbor
warm dust emission, even though only some of the masers are associated
with a detectable ultra-compact (UC) H{\sc II} region
\citep[e.g.][]{Hill05}. This seems to indicate that the masers probe a
range of early phases of massive star formation. The 6.7~GHz masers
are likely pumped by a combination of collisions and emission from
nearby, warm ($T>150$~K) dust, but themselves arise from gas at much
lower temperature ($T<50$~K) with high hydrogen number densities
($n_{\rm H_2} > 10^6$~cm$^{-3}$) and a high methanol abundance
\citep[e.g.][]{Sobolev97, Cragg05}.  Like H$_2$O, methanol is a
non-paramagnetic molecule, and thus both the linear and circular
polarization fractions are small. The first polarization measurements
were made with the Australia Telescope Compact Array (ATCA) on the 6.7
GHz maser toward a handful of southern massive star-forming regions
\citep{Ellingsen02} and linear polarization between few and 10\% was detected.
The first high angular resolution linear polarization maps were
recently made using MERLIN \citep{Vlemmings06c} and the Long Baseline
Array (Dodson 2008). These observations indicate a typical linear
polarization fraction of $\sim 2-3\%$. In addition to the first linear
polarization measurements, a marginal possible detection of circular
polarization, caused by the Zeeman effect, was made for the masers of
the ON1 starforming region \citep{Green07}.

This paper presents the first significant detection of Zeeman
splitting in the 6.7~GHz maser transition for a sample of northern
massive star-forming regions. The observations, data reduction and
error analysis are discussed in \S~\ref{obs} and methanol maser Zeeman
splitting in \S~\ref{circ}. The results are given in \S~\ref{res} and
are discussed in \S~\ref{disc}.

\section{Observations \& analysis}
\label{obs}

\begin{figure*}[ht!]
  \begin{center}
   \resizebox{\hsize}{!}{\includegraphics{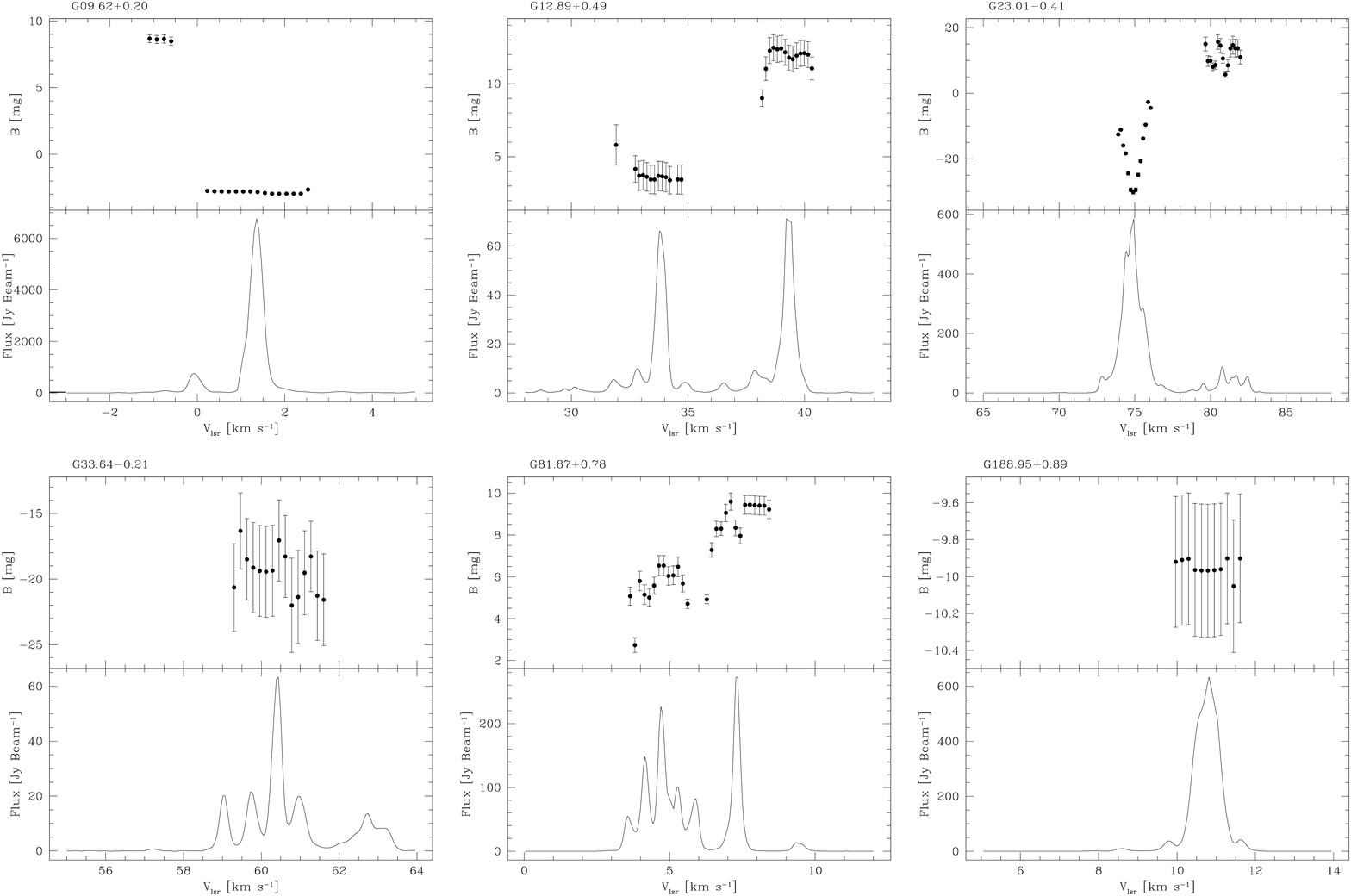}}
  \end{center}
   \hfill \caption{Total intensity spectra (bottom) and magnetic field strengths(top) for six sources of our sample. The magnetic field strength is determined from the measured Zeeman splitting using the current best value for the 6.7~GHz methanol maser splitting coefficient. The Zeeman splitting is derived using the 'running' cross-correlation method (see text).}
   \label{fig_norm}
\end{figure*}

\subsection{Effelsberg observations \& data reduction}

The 6668.512~MHz ($5_1 - 6_0A^+$) methanol maser line of a sample of
massive star-forming regions was observed on Nov 12th 2007 using the
5~cm primary focus receiver of the 100-m Effelsberg\footnote{The 100-m
  telescope at Effelsberg is operated by the Max-Planck-Institut
  f{\"u}r Radioastronomie (MPIfR) on behalf of the
  Max-Planck-Gesellschaft (MPG)} telescope. The data were taken in
position switch mode with a 2 minute cycle time. The full width at
half-maximum (FWHM) beam of the telescope is $\sim120$\arcsec~at the maser
frequency. Data were collected using the fast Fourier transform
spectrometer (FFTS) using two spectral windows, corresponding to the
right- and left-circular polarizations (RCP and LCP). The spectral
windows of $20$~MHz were divided in $16384$ spectral channels,
resulting in a $\sim0.055$~\kms channel spacing and were centered on
the local standard of rest (LSR) source velocities.

The data were reduced using the Continuum and Line Analysis
Single-dish Software (CLASS) package. The amplitudes were calibrated
using scans on 3C123, 3C286 and 3C84. No special efforts were
undertaken to model short time-scale gain fluctuations during the
observations, however, the small gain difference between the RCP and
LCP was corrected. Small RCP and LCP gain differences do not affect
the Zeeman splitting determination when using the cross correlation
method, as with this method the Zeeman splitting measurements are not
reliant on accurate absolute fluxes. It is found that in general the
measured peak fluxes agree well with the literature values in the
methanol maser catalogue of \citet{Pestalozzi05}. From the literature
comparison and from a comparison of the measured flux of G9.62+0.20,
which is the target of regular monitoring, with the flux obtained on
Nov 12th with the Hartebeesthoek radio telescope (S. Goedhart,
priv. communication), the absolute flux errors are estimated to be
$\sim10\%$. Still, for a few sources the measured flux density is up to $50\%$
different from the published values, which is likely due to
variability and/or structural changes in the maser themselves.

\subsection{The source sample}

The sample was taken from the 6.7~GHz methanol maser catalogue by
\citet{Pestalozzi05} and consists of some of the strongest ($>50$~Jy)
northern maser sources observable from Effelsberg. Priority was given to
sources previously observed at high resolution or with existing
OH/H$_2$O maser polarization observations. To detect the Zeeman
splitting the goal was to reach a signal-to-noise of $>3000$ and thus
the total integration time per source was variable. As the
observations were limited to 13~hours, source selection also depended
on observability and was occasionally adjusted based on the flux
detected in a first observing scan. As a result, the noise level is
different for each of the sources and ranges from 20~mJy to 100~mJy
for each polarization.

\subsection{Zeeman splitting determination}
\label{meth}

The circular polarization fraction arising from the small Zeeman
splitting of the paramagnetic methanol molecule, is extremely low
($<0.5\%$). Detection of such small circular polarization fraction is
very sensitive to an accurate relative calibration between the RCP and
LCP signals. Zeeman splitting can be detected relatively
  straightforward using the cross-correlation method, described by
  \citet{Modjaz05}. This method directly determines the Zeeman
  splitting from the RCP and LCP data. It has been shown to be robust
  against the relative RCP and LCP gain calibration errors while still
  reaching a comparable sensitivity as performing S-curve fitting to
  the circular polarization spectrum. Also, essential for single dish
work, it is able to measure Zeeman splitting in the case of spectrally
blended maser features in a straightforward manner without the
  need for fitting a number of Gaussian components to identify
  individual maser features. It was for instance used on Green Bank
Telescope data to determine limits on the Zeeman splitting for two
H$_2$O maser megamaser galaxies \citep{Modjaz05, Vlemmings07}.

As many of the sources will have blended maser features, and to reduce
the effect of remaining RCP and LCP gain calibration errors, the
Zeeman splitting is thus determined using the cross-correlation
method. Applying this method to the entire maser spectrum essentially
means that the resulting Zeeman splitting is a flux weighted average
of the true Zeeman splitting. To examine changes of the magnetic field
over the maser spectrum, one can perform the cross-correlation
analysis on sub-sections of the spectrum. In the case of these
observations, rms noise considerations limit the useful spectral range
over which the analysis can be applied to $>3$~\kms. Thus, to describe
the magnetic field changes over the maser spectrum, cross-correlation
was done in the interval [$(V_i-1.5)$,$(V_i+1.5)$]~\kms, with $V_i$
taken along the spectrum at steps of 3 channel widths. Examples of
such 'running' cross-correlation are shown in Fig.~\ref{fig_norm} with
only magnetic field values plotted that had $>5\sigma$
significance. Special care has to be taken with velocity intervals
which are dominated by the wings of a bright emission peak. In this
case, remaining gain errors between RCP and LCP could create a
spurious Zeeman detection. Although a test with varying gain factors
indicates that the 'running' cross-correlation is robust against cases
with a uniform gain error across the spectrum, small baseline
variations can still cause errors. Thus, only Zeeman determinations in
velocity intervals that included a separate maser feature (i.e. a
local flux maximum) in the inner 80\% of the interval were taken to be
significant. It should be noted that, as the 'running'
  cross-correlation effectively convolves the spectra down to a
  $3$~\kms spectral resolution, apparent gradual magnetic field
  variations are introduced when spectral features with a different
  field strength are blended together. This gradual magnetic field
  change between the maser features is thus not necessarily physically
  the case in the maser region.

\begin{figure*}[ht!]
  \begin{center}
   \resizebox{\hsize}{!}{\includegraphics{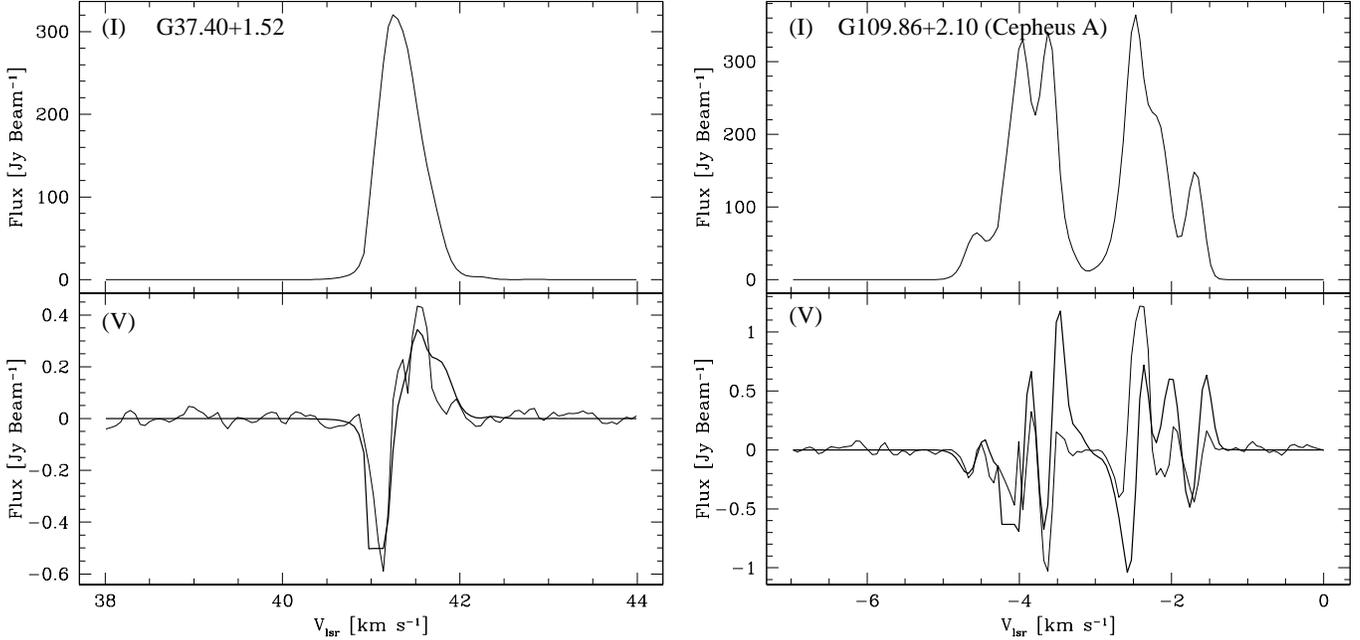}}
  \end{center}
   \hfill \caption{Total intensity and circular polarization spectrum for G37.40+1.52 (left) and G109.86+2.10 (Cepheus A; right). The thick solid line in the bottom panel is best fit fractional total power derivative to the circular polarization spectrum}
   \label{fig_vspec}
\end{figure*}

Although the analysis was performed using the cross-correlation
method, the circular polarization pattern as expected from Zeeman
splitting was detected for several of the least complex sources. The
magnitude of the circular polarization was
$\sim0.2$\%. Fig.~\ref{fig_vspec} shows an example of two sources. The
case of Cepheus~A illustrates the problem with determining magnetic
fields from the circular polarization produced by such a complex
maser. Although the structure of the Cepheus~A circular polarization
is similar to the total power derivative, it is extremely
  difficult to properly distinguish the individual features.

\subsection{Error analysis}

Small amplitude non-Gaussian effects on the spectral baselines can
introduce an additional uncertainty in the Zeeman splitting
determination. To estimate this effect, Monte-Carlo modeling was
performed using artificially generated maser spectra with actual
baselines taken from emission free spectral regions for each of the
sources of the sample. It was found that typically the rms from the
cross-correlation Zeeman splitting calculations need to be increased
by $\sim15\%$ to accommodate this effect. This additional source of
error has been included in the errors quoted below.

\subsection{Beam Squint}

The effect of a slight difference in the pointing center between the
RCP and LCP, as a result of both feeds being off-axis, is called beam
squint \citep[e.g][]{Heiles96}. Beam squint can mimic the effect of
Zeeman splitting when a velocity gradient is present across extended
emission of a spectral line, as the RCP and LCP telescope beam will
then be probing gas at a slightly different velocity. This severely
complicates the Zeeman splitting observations of extended thermal line
emission and needs significant additional effort by rotating the feeds
during the observations. 

However, typically, individual maser features are not very
extended. \citet{Minier02} found that the majority of the methanol
masers consist of a compact core with a diffuse halo structure of up
to a few hundred AU, corresponding to up to 100 mas. Even for the most
extended sources, such as W3(OH), where methanol maser emission
extends over several arcseconds, the bulk of the maser flux which
dominates the Zeeman splitting determination, comes from a region
smaller than 100 mas. The pointing difference between the Effelsberg
RCP and LCP receivers are $\lesssim1\arcsec$
\citep[e.g.][]{Fiebig90}. The effect of beam squint will be most
severe when the velocity gradient exists in the direction of the beam
off-set. Thus, during longer observation scans, any artificial
circular polarization signature will be quenched when the feeds rotate
under the source. However, as the observation scans were occasionally
as short as a few minutes, it was determined to what level beam squint
could contribute to the observed Zeeman splitting. Assuming the worst
case, a velocity gradient across the maser in the direction of the
RCP-LCP beam off-set, a broad maser spectrum of $\sim 5$~\kms and a
maser extension of $100$~mas, the contribution of the beam squint to
the Zeeman splitting is found to be less than $0.002$~m~s$^{-1}$,
corresponding to $0.04$~mG. The beam squint circular polarization
signature and the regular maser Zeeman splitting only becomes similar
if the maser would extend over more than $5$\arcsec.  However, when an
extended maser region consists of individual compact maser features
that dominate the spectrum, the beam squint effect is negligible. It
can thus be concluded that the velocity splitting between RCP and LCP
measured for the 6.7~GHz methanol maser sources is not due to beam
squint.

\section{Methanol maser circular polarization}
\label{circ}

\subsection{Zeeman splitting}

The methanol molecule is non-paramagnetic and as a result the Zeeman
splitting under the influence of a magnetic field is extremely
small. The split energy, $\Delta E_Z$, of an energy level under the
influence of a magnetic field, $B$, is described by $\Delta E_z =
10^{4} g \mu_N M_J B$. Here $M_J$ denotes the magnetic quantum number
for the rotational transition with the rotational quantum number $J$,
$B$ is the magnetic field strength in Gauss, $\mu_N$ is the nuclear
magneton and $g$ is the Land\'e $g$-factor, which thus determines the
magnitude of the Zeeman effect. This factor was investigated many
years ago by \citet{Jen51}, who found empirically that it is probably an
average of the true $g$-factor of several interacting states and can
be described by the equation:
\begin{equation}
g = 0.078 + 1.88/[J(J+1)].
\end{equation}
This is based on laboratory measurements using 25~GHz methanol masers
lines and as a result one has to be careful applying these results to
the 6.7 or 12.2~GHz masers. However, until further laboratory studies,
this remains the best estimate and implies, for the 6.7~GHz methanol
transition, a Zeeman splitting coefficient of
$0.049$~\kms~G$^{-1}$. It is difficult to properly judge the
uncertainty in this value. Conservatively, one expects that magnetic
field strengths determined using this value are within $\sim25$\% of
the true value.

\subsection{Non-Zeeman effects}

There are several, non-instrumental, effects other than Zeeman
splitting that might cause circular polarization of the maser
line. One possibility is a rotation of the axis of symmetry for the
molecular quantum states. This can occur when, as the maser saturates,
the rate for maser stimulated emission $R$ becomes larger than the
Zeeman frequency shift $g\Omega$. While $g\Omega>>R$, the magnetic
field direction is the quantization axis. Then, when $R$ becomes smaller
than $g\Omega$, the molecules interact more strongly with the
radiation field than with the magnetic field and the quantization axis
changes towards the maser propagation direction.  From the Zeeman
splitting coefficient derived above, $g\Omega\approx22B{\rm
  [mG]}$~s$^{-1}$ for the 6.7~GHz methanol maser. The rate for
stimulated emission can be estimated using:
\begin{equation}
R \simeq A k T_{\rm b}\Delta\Omega / 4\pi h\nu.
\end{equation}
Here $A$ is the Einstein coefficient for the maser transition, which
is equal to $0.1532\times10^{-8}$~s$^{-1}$ \citep{Cragg93}, and $k$
and $h$ are the Boltzmann and Planck constants respectively. The maser
frequency is denoted by $\nu$, and $T_{\rm b}$ and $\Delta\Omega$ are
the maser brightness temperature and beaming solid angle. Observations
indicate that $T_{\rm b}\lesssim10^{12}$~K \citep{Minier02}, but
$\Delta\Omega$ is somewhat harder to estimate and decreases rapidly
with increasing maser saturation level. For a typical maser beaming
angle of $\Delta\Omega\approx10^{-2}$, the maser stimulated emission
$R\lesssim4$~s$^{-1}$. Thus, even in the most saturated case
$g\Omega>>R$ when the magnetic field strength is $\gtrsim1$~mG. A
change of quantization axis is thus unable to explain the observed
circular polarization.

Alternatively, \citet{Wiebe98} have shown that the propagation of
strong linear polarization can cause circular polarization when the
direction of the magnetic field changes significantly along the maser
propagation direction. For a smooth change of magnetic field direction
of $\sim1$~rad along the maser, the fractional circular polarization
caused by this effect is approximately $m_l^{2}/4$, where $m_l$ is the
fractional linear polarization. For the typical 6.7~GHz methanol maser
linear polarization fraction of $2-3\%$, this implies that the
generated circular polarization is only $\sim0.02\%$, which is less
than $10\%$ of the observed circular polarization. And, the fairly
constant linear polarization vectors observed at high-resolution
\citep[e.g.][]{Vlemmings06c}, indicate that the magnetic field
rotation along the maser path length is likely much smaller than
$1$~rad, making the propagation of linear polarization an unlikely
source of the methanol maser circular polarization.

Finally, a velocity gradient in intrinsic circular polarization
(i.e. not caused by the magnetic field) could mimic the effect of
Zeeman splitting. However, as in a maser the stimulated emission
quickly dominates the intrinsic emission, any intrinsic circular
polarization quickly becomes negligible. This is described in more
detail in e.g. \citet{Elitzur98}.

\section{Results}
\label{res}
\begin{table*}
\caption{Zeeman splitting results}
\begin{tabular}{|l|l|c|c|c|c|c|c||c|}
\hline
{Source} & & $\alpha_{\rm J2000}$ & $\delta_{\rm J2000}$ & $V_{\rm LSR}$ & Peak Flux & Int. Flux & $\Delta V_z$ & $|B|$ \\
 & & $hh~mm~ss$ & $^\circ~'~''$ & \kms & (Jy/beam) & (Jy/beam) & m~s$^{-1}$ & (mG) \\
\hline
\hline
09.62+0.20$^a$ &	              & 18~06~14.66 &	-20~31~31.60 & 1.0  & 6757 & 3091 & $-0.137\pm0.002	$ & $-2.80\pm0.04 $ \\	 
 &	              &  &	 & -0.1$^b$  &  &  & $0.42\pm0.02	$ & $8.6\pm0.4 $ \\	 
 &	              &  &	 & 1.2$^b$  &  &  & $-0.145\pm0.002	$ & $-2.96\pm0.04 $ \\	 
12.89+0.49$^a$ & IRAS 18089-1732 & 18~11~51.46 &  	-17~31~28.84 & 39.3 & 71   & 105   & $ 0.41\pm0.03   $ & $  8.4\pm0.7  $ \\
 &	              &  &	 & 34.8$^b$  &  &  & $0.18\pm0.05	$ & $4\pm1 $ \\	 
 &	              &  &	 & 39.2$^b$  &  &  & $0.59\pm0.05	$ & $12\pm1 $ \\	 
23.01-0.41$^a$ &  	              & 18~34~40.37 &  	-09~00~38.30 & 74.8 & 585  & 1016  & $-1.49\pm0.03  	$ & $-30.3\pm0.7  $ \\
 &	              &  &	 & 74.9$^b$  &  &  & $-1.49\pm0.02	$ & $-30.3\pm0.4 $ \\	 
 &	              &  &	 & 81.0$^b$  &  &  & $0.7\pm0.1	$ & $14\pm2 $ \\	 
25.71+0.04 &  	              & 18~38~03.10 &  	-06~24~32.00 & 95.6 & 590  & 591  & $ <0.06	        $ & $ <1.2        $ \\
25.83-0.18 &  	              & 18~39~04.70 &  	-06~24~17.00 & 90.7 & 69   & 84   &	$ 0.99\pm0.20	$ & $   20\pm4    $ \\
28.15+0.00 &  	              & 18~42~41.00 &  	-04~15~21.00 & 101.3& 30   & 24   &	$ <1.8          $ & $ <37	  $ \\
31.28+0.06 & IRAS 18456-0129 & 18~48~12.38 &  	-01~26~22.60 & 110.4& 74   & 174  & $ 2.06\pm0.26   $ & $ 	42\pm5    $ \\
32.03+0.06 & IRAS 18470-0050 & 18~49~37.04 &  	-00~46~50.10 & 98.7 & 69   & 81   & $ <0.51 	$ & $ <11         $ \\
33.64-0.21 &   	      & 18~53~28.70 &  	+00~31~58.00 & 58.6 & 63   & 54   &	$-0.89\pm0.17	$ & $  -18\pm3    $ \\
35.20-0.74 & IRAS 18556+0136 &	18~58~12.98 &  	+01~40~37.50 & 30.5 & 169  & 167  &	$ 0.81\pm0.04  	$ & $ 16.5\pm0.7  $ \\
35.20-1.74 & W48   	      &	19~01~45.60 &  	+01~13~28.00 & 41.5 & 476  & 643 &	$ 0.32\pm0.02	$ & $  6.4\pm0.3  $ \\
37.40+1.52 & IRAS 18517+0437 &	18~54~13.80 &  	+04~41~32.00 & 41.0 & 320  & 193  &	$ 0.75\pm0.02   $ & $ 15.4\pm0.4  $ \\
43.80-0.13 & W49N            & 19~11~55.10 &   +09 36 00.00 & 40.0 & 35   & 81   &	$<1.9 	        $ & $ <39         $ \\
49.49-0.39 & W51-e1/e2             & 19~23~44.50 &   +14~30~31.00 & 59.0 & 1029  & 885  & $ 0.72\pm0.04	$ & $ 14.7\pm0.8  $ \\
69.52-0.97 & ON1	      &	20~10~09.07 &	+31~31~34.40 & 11.6 & 96   & 45   &	$ <0.2	        $ & $ <4.1        $ \\
78.10+3.64 & IRAS 20126+4104 & 20~14~26.04 &   +41~13~33.39 & -6.1 & 60   & 77   &	$ <0.8	        $ & $ <17         $ \\
% &	              &  &	 & -8.6$^b$  &  &  & $-1.1\pm0.3	$ & $-22\pm6 $ \\	 
% &	              &  &	 & -5.0$^b$  &  &  & $0.24\pm0.05	$ & $5\pm1 $ \\	 
81.87+0.78$^a$ & W75N	      &	20~38~36.80 &	+42~37~59.00 & 5.0  & 273  & 317  &	$ 0.40\pm0.02	$ & $  8.2\pm0.3  $ \\
 &	              &  &	 & 5.0$^b$  &  &  & $0.28\pm0.02	$ & $5.7\pm0.4 $ \\	 
 &	              &  &	 & 7.3$^b$  &  &  & $0.46\pm0.02	$ & $9.4\pm0.4 $ \\	 
109.86+2.10 & Cepheus A         &	22~56~18.09 &  	+62~01~49.45 & -4.2 & 364  & 484  &	$ 0.39\pm0.01	$ & $  8.1\pm0.2  $ \\
111.53+0.76$^a$ & NGC~7538        &	23~13~45.36 &  	+61~28~10.55 & -56.1& 233  & 514  &	$ 0.79\pm0.03 	$ & $ 16.2\pm0.6  $ \\
 &	              &  &	 & -58$^b$  &  &  & $0.78\pm0.04	$ & $16.0\pm0.8 $ \\	 
 &	              &  &	 & -56$^b$  &  &  & $0.59\pm0.02	$ & $12.1\pm0.4 $ \\	 
133.94+1.04 & W3(OH)         &	02~27~04.72 &  	+61~52~24.73 & -44.0& 3705 & 8198 & $ 0.141\pm0.003	$ & $ 2.87\pm0.05 $ \\
173.49+2.42 & S231	      &	05~39~13.06 &	+35~45~51.29 & -13.0& 52   & 53   &	$ 0.95\pm0.11	$ & $   19\pm2    $ \\
174.19-0.09 & AFGL~5142       & 05~30~42.00 &   +33~47~14.00 & 2.1  & 55   & 34   &	$ < 0.8   $ & $	<18    $ \\
188.95+0.89 & IRAS 06058+2138 &	06~08~53.35 &  	+21~38~28.67 & 10.9 & 633  & 485  &	$-0.49\pm0.02 	$ & $ -9.9\pm0.4  $ \\
192.60-0.05 & S255           & 06~12~54.02 &   +17~59~23.00 & 5.0  & 94   & 78   &	$ 0.47\pm0.05	$ & $  9.6\pm0.9  $ \\
\hline
\multicolumn{9}{l}{$^a$ Significant magnetic field changes across the maser spectrum.}\\
\multicolumn{9}{l}{$^b$ Velocity of the maser feature for which Zeeman splitting was determined separately.}\\
\end{tabular}
\label{zeeman_res}
\end{table*}

The results of the Zeeman splitting analysis using the RCP-LCP
cross-correlation method are presented in
Table~\ref{zeeman_res}. The table lists the source name, position,
central $V_{\rm LSR}$ velocity, peak and integrated flux and the
measured Zeeman splitting $\Delta V_z$. As described in \S~\ref{meth},
this value corresponds to the flux averaged Zeeman splitting of the
entire maser spectrum. In the final column it lists the strength of
the magnetic field component along the line-of-sight ($B_{||}$)
determined using the current best value of the 6.7~GHz methanol maser
Zeeman splitting coefficient as discussed in \S~\ref{circ}. Throughout
the paper, the quoted magnetic field strength will correspond to this
value. Should in the future a better value of the Zeeman coefficient
be determined then all magnetic field values need to be adjusted
correspondingly.

\begin{figure*}
 \centering
 \resizebox{0.9\hsize}{!}{\includegraphics{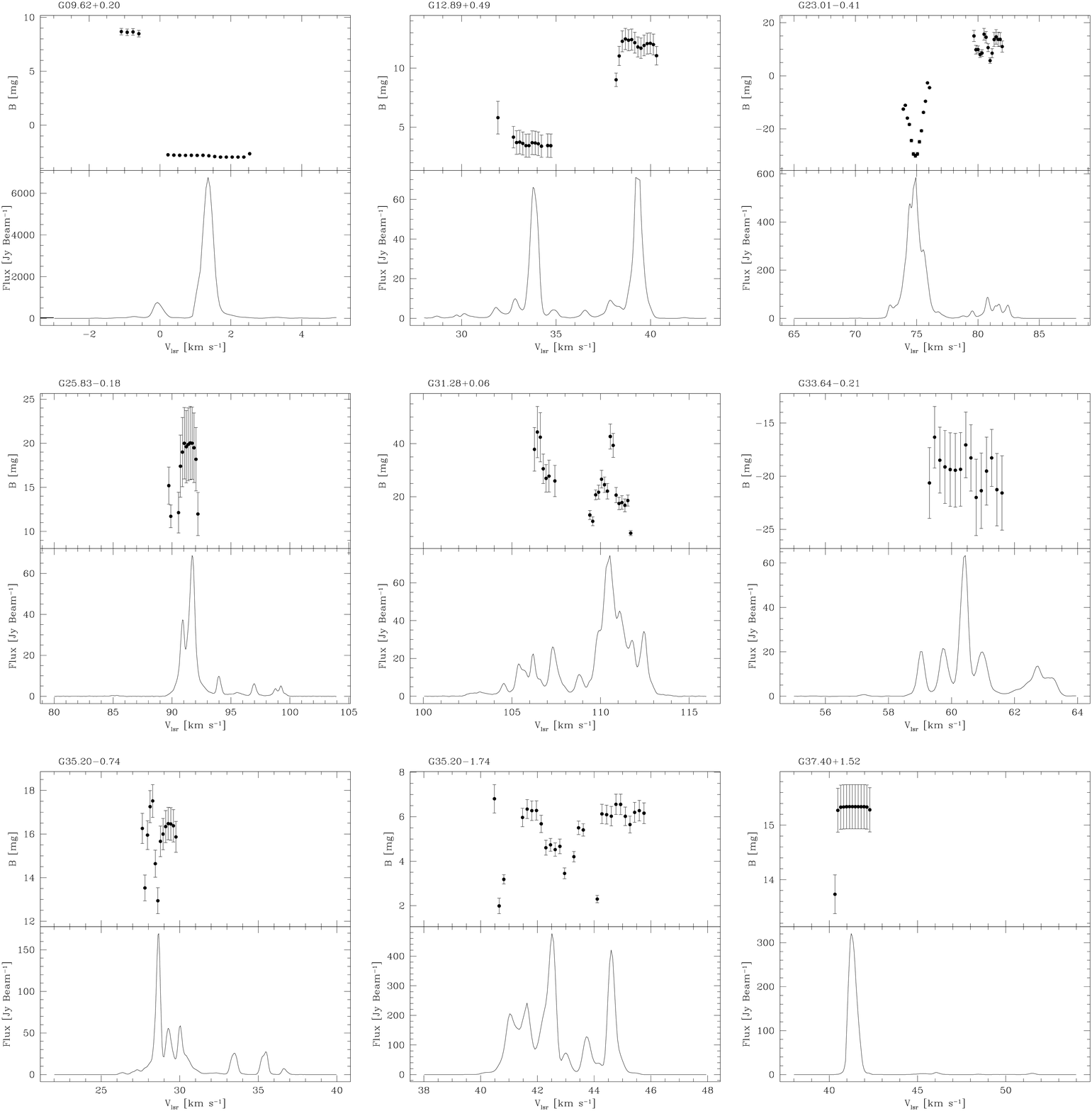}}
 \caption{Total intensity spectra (bottom) and magnetic field strength (top) for all the sources of our sample with significant Zeeman splitting detection. The magnetic field strength is determined from the measured Zeeman splitting using the current best value for the 6.7~GHz methanol maser splitting coefficient. The Zeeman splitting is derived using the 'running' cross-correlation method (see \S~\ref{meth}).}\label{specfig1}
 \end{figure*}

\begin{figure*}
 \centering
 \resizebox{0.9\hsize}{!}{\includegraphics{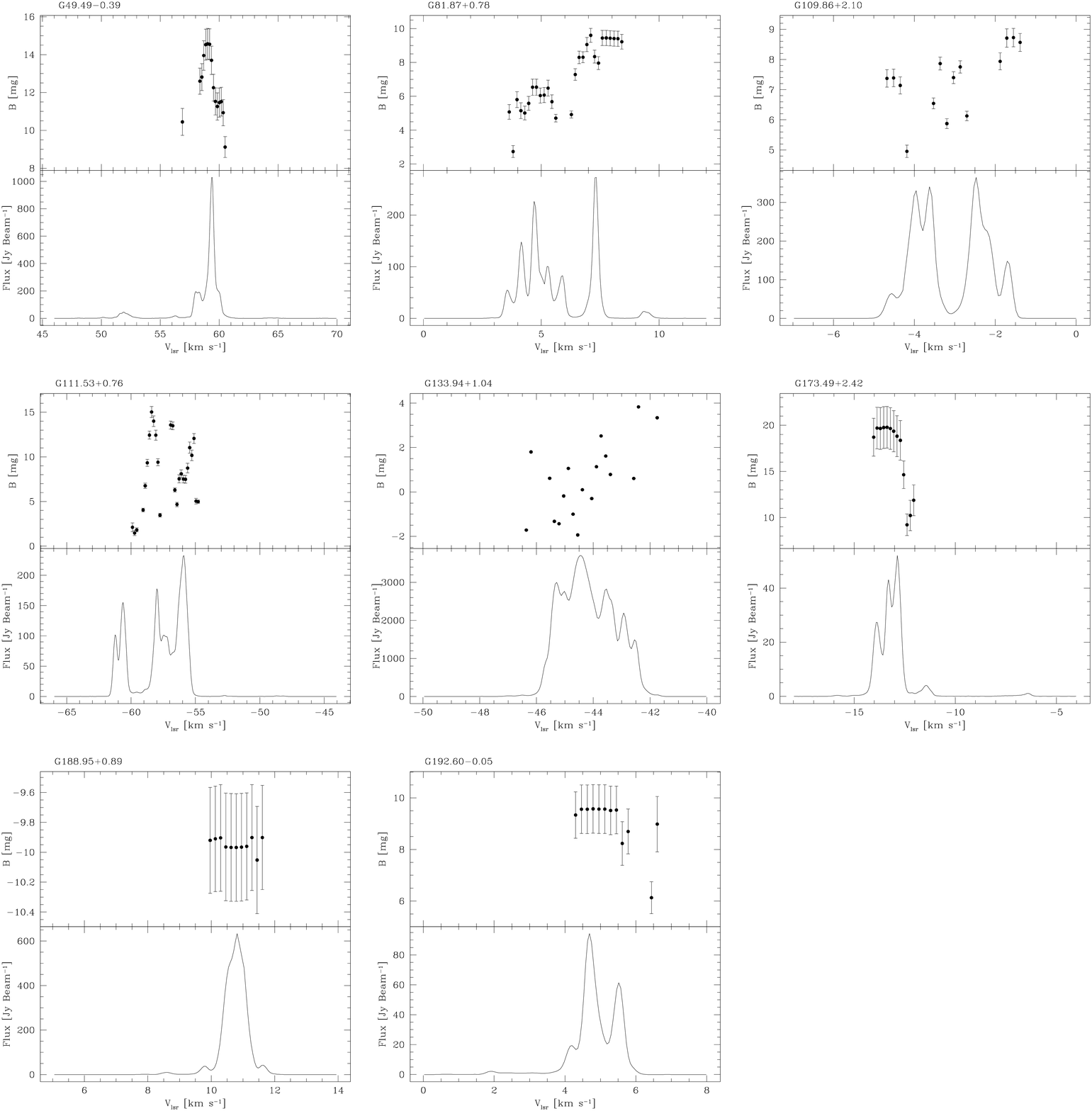}}
 \caption{As Fig.~\ref{specfig1}.}\label{specfig2}
 \end{figure*}

\begin{figure*}
 \centering
 \resizebox{0.9\hsize}{!}{\includegraphics{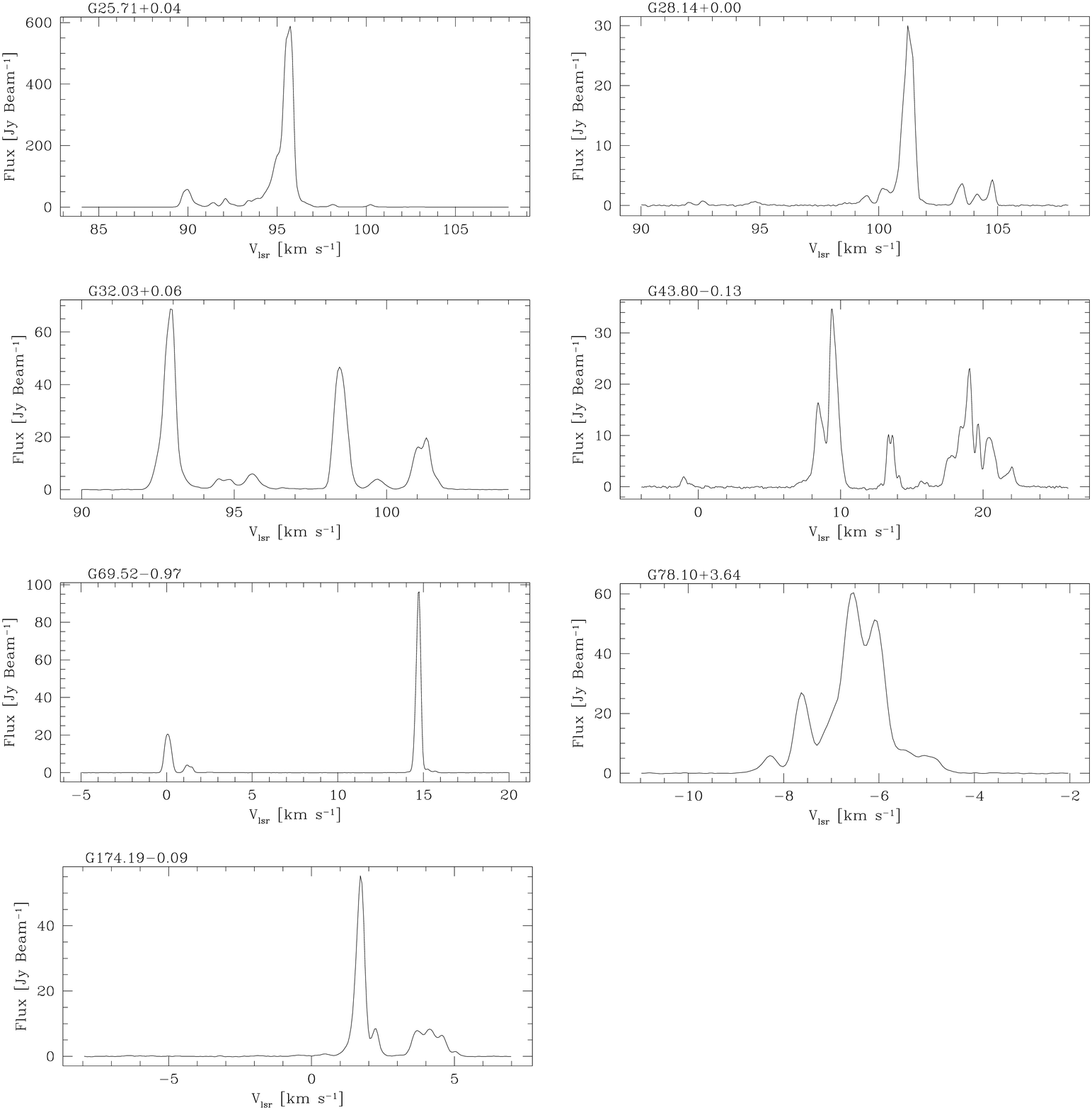}}
 \caption{Total intensity spectra of the sources in our sample for which no significant Zeeman splitting was detected.}\label{specfig3}
 \end{figure*}

Fig.~\ref{fig_norm} shows six methanol maser spectra and
corresponding magnetic fields derived using the running
cross-correlation method described in \S~\ref{meth}. The figure shows 
that for a number of the sources, the magnetic field changes across
the maser. For those sources, Table~\ref{zeeman_res} also lists
separately the magnetic field derived for sub-sections of the
spectrum. Total intensity spectra and magnetic field strengths for all the sources in the sample with a significant Zeeman detection
are presented in the Online material as Figs.~\ref{specfig1} and \ref{specfig2}. Online Fig.~\ref{specfig3} shows the total intensity spectra for the sources without Zeeman detection.

\subsection{Individual Sources}

A number of the maser sources from the sample have been observed at high
angular resolution, providing information on the morphology of the
methanol maser region. Additionally, several sources have been the
target of magnetic field measurements using mainly OH and/or H$_2$O
maser polarization observations. Sources for which
  high-resolution 6.7~GHz observations and/or additional magnetic
  field measurements are available are discussed here in detail.

\subsubsection{09.62+0.20}

The 6.7 and 12.2~GHz methanol masers associated with the high-mass
star formation complex G09.62+0.20 undergo periodic flares with a
period of $244$~days, making these the first reported incidence of
periodic variations in a high-mass star-forming region
\citep{Goedhart03, Goedhart04}. The origin of the periodic behavior,
however, is still unclear. The 6.7~GHz methanol masers have been
mapped with the ATCA by \citet{Phillips98}, and the strongest features
are shown to be associated with the hypercompact H{\sc II} region
labeled E by \citet{Garay93}. This region also hosts OH masers for
which Zeeman splitting observations indicates a magnetic field
strength of $\sim5$~mG \citep{Fish05}. As seen in Fig.\ref{fig_norm},
the magnetic field determined from the 6.7~GHz methanol maser Zeeman
splitting reverses direction between the second-strongest maser
feature at $V_{\rm lsr}=-0.1$~\kms ($B_{||}=8.6$~mG) and the strongest
feature at $1.2$~\kms ($B_{||}=-2.96$~mG), even though the ATCA image
shows that the features are located within $\sim100$~mas of each
other. The periodicity of the flaring of both maser features is the
same, with the weaker of the two features flaring with a $\sim1$~month
delay. There is an intriguing possibility that the maser flaring and
time delay is the result of a periodic change in the magnetic field
orientation due to a form of magnetic beaming
\citep[e.g.][]{Gray94}. However, this hypothesis needs to be tested by
magnetic field monitoring observations and improved maser modeling.

\subsubsection{12.89+0.49, IRAS 18089-1732}

This young massive star-forming region contains methanol, H$_2$O and
OH maser emission as well as a weak hypercompact H{\sc II} region. The
6.7~GHz methanol masers have been mapped with the ATCA and consist of two
regions separated by $\sim1.5$\arcsec \citep{Walsh97}. The strong maser feature at $V_{\rm
  LSR}=33.6$~\kms, which shows the smallest Zeeman splitting
($B_{||}=4$~mG) is located closest to the sub-millimeter continuum
peak imaged by \citet[e.g.][]{Beuther05} at high-resolution with the
submillimeter array (SMA). The higher velocity features, with
$B_{||}=12$~mG, are located in the direction of the North-South
molecular outflow found in the same SMA observations. The OH masers on
the other hand, are located in the direction of extended molecular
emission in the East-West direction, which shows indications of a
rotation signature perpendicular to the outflow \citep{Beuther05, Beuther08}. The
magnetic field measured from OH maser Zeeman splitting is $B=3.93$~mG
\citep{Ruiz06}.

\subsubsection{23.01-0.41}

The Zeeman splitting for this source, and consequently the magnetic
field, shows a very strong negative magnetic field for the brightest
maser peak at $V_{\rm LSR}=74.9$~\kms ($B_{||}=-30.3$~mG), while the
other maser features have a positive field direction with a strength
of $B_{||}\approx 14$~mG. In Fig.~\ref{fig_norm}, this results in a
dip towards an increasingly negative magnetic field across the maser,
because, as described in \S~\ref{meth}, the running average calculates
the Zeeman splitting over a $3$~\kms interval. \citet{Szymczak04} find
a 1667~MHz OH maser feature with a magnetic field strength of
$B=-0.2$~mG at $V_{\rm LSR}=74.31$~\kms. Unfortunately, no high
resolution maps are available to check the morphology of the maser
source.

\subsubsection{31.28+0.06, IRAS 18456-0129}

The largest Zeeman splitting was measured for the masers of the UC
H{\sc II} region G31.28+006, which is part of the giant H{\sc II}
region W43, and correspond to $B_{||}=42$~mG. The 6.7~GHz methanol
masers were observed with the European VLBI Network (EVN) and reveal a
complex distribution over $\sim400\times400$~mas \citep{Minier00}. No
other maser polarization observations are found in the literature.

\subsubsection{33.64-0.21}

As seen in Fig.~\ref{fig_norm}, the magnetic field of this source is
stable over the entire velocity range with a flux averaged strength of
$B_{||}=-18$~mG. The only other magnetic field measurement for
33.64-0.21 comes from a single 1720~mG OH maser feature at $V_{\rm
  LSR}=60.23$~\kms, which has $B=-1$~mG \citep{Szymczak04}.

\subsubsection{35.20-0.74, IRAS 18556+0136}

The methanol masers of the bipolar outflow source G35.20-074N have
only recently been mapped using the Japanese VLBI Network
\citep{Sugiyama08}, showing that the brightest features make up two
compact regions separated by more than 2\arcsec in a direction
perpendicular to the CO outflow, along a putative molecular disc
\citep{Dent85}. The Zeeman splitting observations indicate a stable
flux averaged magnetic field strength of $B_{||}=16.5$~mG, which is
likely dominated by the strongest maser feature at $V_{\rm
  LSR}=28.7$~\kms in the southern maser cluster. The OH maser
polarization of this source has been studied with MERLIN by
\citet{Hutawarakorn99}, who find that the magnetic field reverses on
opposite sides of the disc. The OH maser distribution is thought to
lie along the disc, with the southern masers tracing a mean field
$B\approx4$~mG and a northern maser feature having $B=-2.5$~mG.

\subsubsection{35.20-1.74, W48}

The UC H{\sc II} region G 35.20-1.74, in the Galactic H{\sc II} region W48,
is believed to be a site of massive star formation and has both OH and
methanol masers. EVN observations of the
6.7~GHz methanol maser region shows a ring-like structure of $\sim
200\times400$~mas \citep{Minier00}. Although the spectrum is
complex, the Zeeman splitting observations do not show any large
variations and yield a flux averaged field strength of
$B_{||}=6.4$~mG. The magnetic field at the periphery of the UC H{\sc
  II} region has been determined from carbon recombination line
observations, which give a field of $B=2.9$~mG at a hydrogen number
density of $n_{\rm H_2}=8.5~10^6$~cm$^{-3}$ \citep{Roshi05}.

\subsubsection{49.49-0.39, W51-e1/e2}

W51 is one of the most luminous massive star formation complexes of
our Galaxy and can be divided in three regions, W51A, W51B and W51C
\citep{Carpenter98}. The maser site G49.49-0.39 is associated with
W51A and particularly with W51-e1/e2 \citep{Caswell95}. OH maser
polarization observations reveal a total of 46 Zeeman pairs near e1
and e2, with the region having predominantly a positive magnetic field
direction which reverses in the northern part of e1
\citep{Fish06}. W51e2 has two Zeeman pairs that imply a magnetic field
of the order of $\sim20$~mG \citep{Argon02}, with the rest of the OH
masers indicating $B\sim5$~mG. For the 6.7~GHz methanol masers the
Zeeman splitting indicates a flux averaged $B_{||}=14.7$~mG.

\subsubsection{69.52-0.97, ON1}

The 6.7~GHz methanol maser emission and polarization was recently
mapped with MERLIN by \citet{Green07}. Those observations indicate
a tentative first detection of 6.7~GHz methanol maser Zeeman splitting
of $9\pm3$~m~s$^{-1}$, corresponding to a field strength of
$-18$~mG. This marginal detection could not be confirmed in the
observations presented here. However, the Zeeman splitting measured by
Green et al. was found on a maser feature that, in lower resolution
observations, would be blended both positionally and spectroscopically
with the brightest maser. As a results, the flux averaged Zeeman
splitting measurement is biased toward a possibly negligible
magnetic field of the brightest feature.

\subsubsection{81.87+0.78, W75N}

The 6.7~GHz methanol masers of the very active region of massive star
formation W75N have been mapped with the EVN \citep{Minier00}. Their
map reveals that the masers make up two distinct regions, an elongated
region of $\sim200$~mas with $V_{\rm LSR}<6$~\kms and a compact
feature $\sim500$~mas South-East at $V_{\rm LSR}=6.8$~\kms. The
observations shown in Fig.~\ref{fig_norm} highlight the distinct
nature of these maser features as the masers in the linearly extended
structure have $B_{||}=5.7$~mG while those in the compact region have
$B_{||}=9.5$~mG. W75N has been the target of numerous OH maser
polarization observations indicating another possible disc related
field reversal and typical magnetic field strengths of
$|B|\approx5-7$~mG \citep[e.g.][]{Hutawarakorn02, Slysh02}. Magnetic
field measurements during a OH maser flare in W75N reveal the
strongest OH maser magnetic field $B=40$~mG to date \citep{Slysh06}.

\subsubsection{109.86+2.10, Cepheus A}

\begin{figure}[t!]
  \begin{center}
   \resizebox{\hsize}{!}{\includegraphics{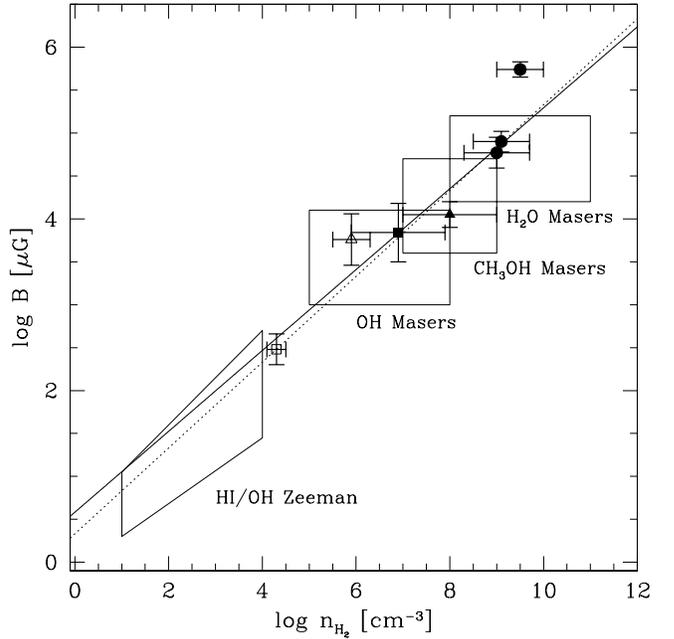}}
  \end{center}
  \hfill \caption{Magnetic field strength $B$ in the massive star
    forming region Cepheus A measured from Zeeman measurements as a
    function of $n_{\rm H_2}$, the number density of neutral hydrogen.
    The solid triangle indicates the methanol maser measurement from
    the Effelsberg observations, the number density is taken to
      be in the range typical of 6.7~GHz methanol masers
      \citep{Cragg05}, but cannot be determined precisely without
      observations of additional transition lines. Also indicated are
    the H$_2$O maser measurements \citep[solid dots;][]{Vlemmings06},
    average OH maser measurement \citep[solid
    square;][]{Bartkiewicz05}, NH$_3$ measurement \citep[open
    square;][]{Garay96} and recent dust polarization measurement
    \citep[open triangle;][]{Curran07}. The error bars on the
      density of these data are taken from the respective
      publications. The solid line is the theoretical relation
    $B\propto n^{0.5}$ fixed to the OH maser measurement. The dashed
    line is the error weighted best fit line to the presented
    measurements (the individual H$_2$O maser measurements were
    averaged). Additionally, the boxes indicate the range of
    literature values for HI/non-masing OH, OH maser, methanol maser
    (this paper) and H$_2$O maser Zeeman splitting observations for
    other massive star formation regions.}
   \label{fig_cepha}
\end{figure}

The masers of the Cepheus A massive star formation region has been
studied in great detail \citep[e.g.][and references therein]{Vlemmings06}. The 6.7~GHz methanol maser distribution has been mapped with
the Japanese VLBI Network \citep{Sugiyama08} and the EVN
(Torstensson et al., in prep.). These observations show that the
masers are found in an elongated structure of $\sim1500$~mas across
Cepheus~A~HW2. The measured methanol maser magnetic field is stable at
$B_{||}=8.1$~mG. 

Cepheus A is the source in the sample for which the most magnetic
field measurements at different hydrogen density are
available. Fig.~\ref{fig_cepha} presents on overview of all these
measurements. It seems clear that the magnetic field strength $B$ in
Cepheus A has a power-law dependence on the hydrogen number density
$n_{H_2}$, with the best-fit giving $B\propto n^{-0.47\pm0.08}$. This
is consistent with the empirical relation $B\propto n^{-0.47}$ from
low-density molecular cloud Zeeman observations by
\citet{Crutcher99}. Of course, one has to be cautious when relating
the maser observations with the other Zeeman splitting results and the
dust polarization observations at very different scales. Additionally,
the methanol and H$_2$O maser magnetic field strengths depend on an
assumption with regard to the angle between the maser propagation
direction and the magnetic field and their number density has to
  be determined using maser excitation models.

\subsubsection{111.53+0.76, NGC~7538}

The methanol masers of NGC~7538 have been proposed to trace a disk
around a high-mass protostar \citep{Pestalozzi04}, however, this
interpretation has been questioned by \citet{DeBuizer05} who find that
the maser might be related to an outflow. The observations presented
here indicate a flux averaged magnetic field $B_{||}=16.2$~mG.   The 'running' cross-correlation however, reveals a more complex
  picture. The spectrum is made up of several strong maser features
  with apparently different magnetic field strengths. The strongest
  magnetic field $\sim16$~mG is measured on the maser feature at
  $V_{\rm LSR}=-58$~\kms, while $\sim12$~mG is found for the maser at
  $V_{\rm LSR}=-56$~\kms. In both cases the 'running'
  cross-correlation derived magnetic field is somewhat supressed as
  the masers between $V_{\rm LSR}=-58$ and $-56$~\kms~ have a field of
  only a few mG. Meanwhile, no significant magnetic field is detected
  in the masers near $V_{\rm LSR}=-61$~\kms, implying a field strength
  $B_{\rm ||}\lesssim5$~mG. The flux averaged magnetic field is
  marginally larger than the field determined using the 'running'
  cross-correlation for individual maser features. This is due to the
  fact that, in determining the flux averaged field strength, strong
  fields, which have nevertheless $<5\sigma$ significance in the
  smaller $3$~\kms~intervals used for the 'running' cross-correlation,
  still contribute. OH maser measurements indicate the region
undergoes a field reversal and has $|B|=1$~mG \citep{Fish06}.

\subsubsection{133.94+1.04, W3(OH)}

Linear polarization of the 6.7~GHz methanol masers of W3(OH) is
described in \citet{Vlemmings06c}, who also give a compilation of
previous OH maser magnetic field strength measurements. The
polarization observations show that the magnetic field traces the
extended methanol filament but has a more complex structure in the
dominating compact region described in \citet{HS06}. The Zeeman
splitting presented here corresponds to a flux averaged magnetic field
strength of $B_{||}=2.87$~mG. However, due to the rich spectrum and
large extent of W3(OH) a direct comparison with other Zeeman splitting
measurements is impossible and would need high resolution
observations.

\section{Discussion}
\label{disc}
\subsection{Magnetic field strength}

Significant Zeeman splitting was detected for 17 out of 24 sources and
indicates an absolute magnetic field strength component $|B_{||}|$
along the maser propagation direction between $2.8$ and $42$~mG. There
are several effects that bias the observed field strength towards
higher or lower values. As is the case for H$_2$O masers, low spatial
resolution observations, velocity gradients across
the maser and increased maser saturation tend to cause an
underestimate of the magnetic field strength \citep{Sarma01,
  Vlemmings06b}. As discussed above, beam squint can create a false
Zeeman splitting signature, however, for milliarcsecond-scale masers
this is unlikely to contribute more than $\sim0.04$~mG.  Finally, the
Zeeman coefficient which is used to determine the magnetic field is
uncertain, as is described in \S~\ref{circ}, which leads to a
systematic bias in the magnetic field. Thus the uncertainty in the measured
absolute magnetic field strength will be substantial, with most of the
described effects biasing it towards a value that is lower than the
actual field strength. Weighing the field strength by measurement
significance, the average magnetic field $\langle B_{||,\rm
  meth}\rangle=12$~mG. This needs to be corrected for a random angle between
magnetic field and the line-of-sight, which implies for the absolute field strength $|B|=2\langle B_{||}\rangle$ \citep[e.g.][]{Crutcher99}. This
gives $|B_{\rm meth}|=23\pm6$~mG. The error on the
absolute field strength is dominated by the estimated uncertainty in the Zeeman
coefficient.  This field strength is larger than that found in
the OH maser regions, with the average OH maser determined field strength
being $\sim4$~mG \citep{Fish06}.
 
The dynamical importance of the magnetic field can be quantified by
defining a critical magnetic field strength $B_{\rm crit}=(8\pi\rho
v^2)^{1/2}$ for which the dynamic and magnetic pressure are
equal. Here $\rho$ and $v$ are the density and velocity of the maser
medium respectively. \citet{Cragg05} find that the current
  observational limits suggest that the majority of the 6.7~GHz maser
  sources occur near the high-density limit of the maser range
  ($n_{\rm H_2}=10^7-10^9$~cm$^{-3}$), with a typical value of $n_{\rm
    H_2}=10^8$~cm$^{-3}$. Taking this density and a typical gas
  velocity of $\lesssim5$~\kms~from proper motion measurements
  \citep[e.g.][]{Xu06}, $B_{\rm crit}\approx12$~mG. The measured
  magnetic field vales are thus comparable to $B_{\rm crit}$ and hence
  dynamically important.

\subsection{Comparison with OH masers}
\begin{figure}[t!]
  \begin{center}
   \resizebox{\hsize}{!}{\includegraphics{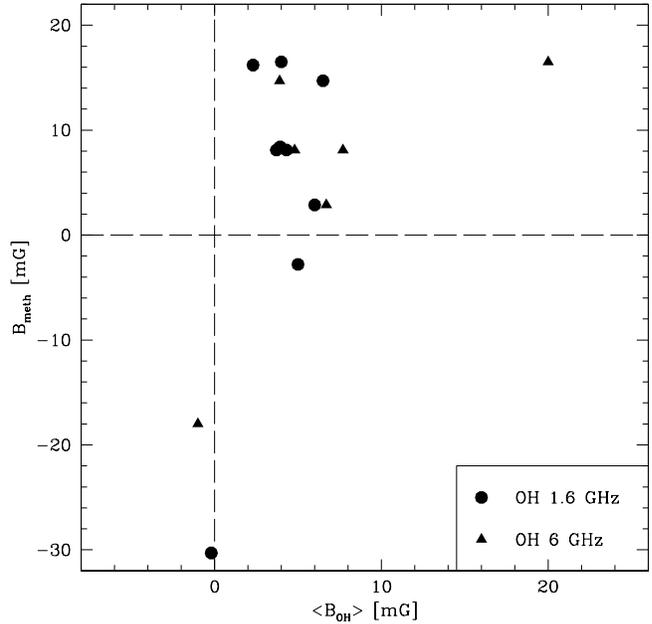}}
  \end{center}
  \hfill \caption{Measured flux averaged 6.7~GHz methanol maser
    magnetic field strength vs. average 1.6 and 6~GHz OH maser magnetic
    field strengths for those sources where both are available. Most literature
    values were taken from the compilation of molecular cloud magnetic
    field measurements by \citet{Han07}.}
   \label{fig_ohmeth}
\end{figure}

Theoretical modeling indicates that, given sufficient abundance, both
1.6 and 6~GHz OH masers as well as the methanol masers can be pumped
simultaneously \citep{Cragg02}. However, observations have shown that,
although they are closely associated, co-propagation between OH and
methanol masers is rare \citep[e.g.][]{Etoka05}. Still, as the masers
are closely related, one might expect a relation between the OH
magnetic field strengths and those derived from the methanol
masers. As seen in Fig.~\ref{fig_ohmeth}, no clear relation is found
for those sources which have magnetic field strength measurements from
both maser species. This is likely due to the fact that the flux
averaged methanol maser magnetic field strengths cannot be easily
related to the OH maser measurements often performed at much higher
resolution.

\begin{figure}[t!]
  \begin{center}
   \resizebox{\hsize}{!}{\includegraphics{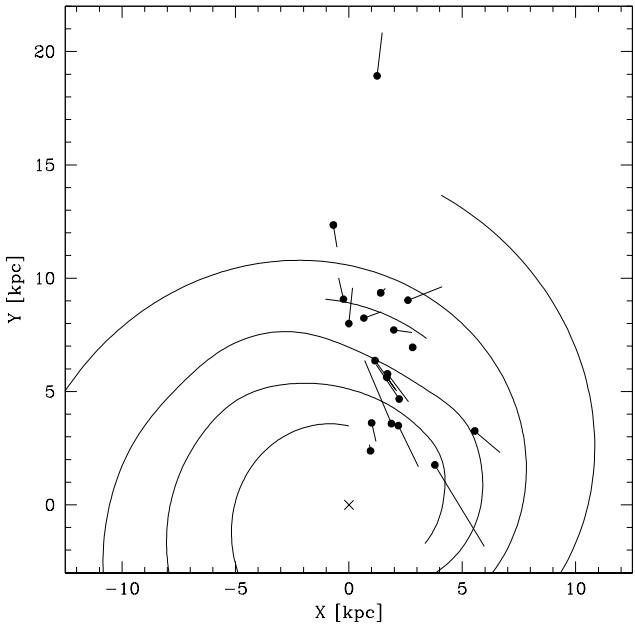}}
  \end{center}
  \hfill \caption{Magnetic field direction and strength along the
    line-of-sight derived from the methanol maser Zeeman splitting
    observations presented in this paper projected onto the Galactic
    plane. The solid dots are the observed star-forming region with
    kinematic distances from \citet{Pestalozzi05}. These are known to
    be highly uncertain. The length of the vectors is scaled with
    $B_{||}$. The approximate location of the spiral arms is indicated
    as taken from \citet{Taylor93}.}
   \label{fig_gal}
\end{figure}

Fig.~\ref{fig_ohmeth} does, however, show that the magnetic field
directions determined from methanol and OH masers are fully
consistent. As discussed for some of the individual sources, this even
holds for regions where the OH masers show a magnetic field reversal.
This implies that the magnetic field orientation derived from the
methanol maser regions, as that from OH masers, could be indicative of
the Galactic magnetic field. This is illustrated in
Fig.~\ref{fig_gal}, although a much larger number of measurements as
well as accurate distances to the star-forming regions would be
needed. Since several efforts are underway to determine accurate
distances to methanol maser regions \citep[e.g][]{Xu06}, methanol maser
Zeeman splitting measurements will provide an important opportunity to
study the Galactic magnetic field \citep[e.g.][]{Han07}.

\subsection{The magnetic field vs. density relation}

Fig.~\ref{fig_cepha} shows the $B$-field measurements of both maser
and non-maser Zeeman observations as a function of hydrogen number
density. The figure seems to indicate that the $B$-field follows the
$B\propto n^{0.5}$ density scaling law discussed in
\citet[e.g][]{Mouschovias99} over an enormous range of densities. This makes
the maser measurements consistent with low-density molecular cloud
measurements \citep{Crutcher99}, implying that the magnetic field
remains partly coupled to the gas up to the highest number
densities. However, some care has to be taken in interpreting this
relation, as for instance the shock excited masers are short-lived
(H$_2$O masers have a typical lifetime $\tau_m \sim 10^8$~s) compared
to the typical ambipolar diffusion time-scale at the highest densities
($\tau_d \sim 10^9$~s). This suggests that in the non-masing gas of
similar densities, magnetic field strengths are likely lower due to
ambipolar diffusion. Still, the maser magnetic field measurements
strongly imply a dynamical importance of magnetic fields during the
high-mass star formation process, with the methanol maser observations
presented here filling the gap between the OH and H$_2$O maser
observations.

\section{Concluding remarks}
\label{concl}
This paper presents the first significant Zeeman splitting
measurements obtained on the 6.7~GHz methanol maser. As this highly
abundant maser uniquely pinpoints massive star formation, a detection
of the Zeeman splitting gives the opportunity to measure the magnetic
field in a large number of high-mass star-forming regions at densities
of $n_{H_2}\approx10^8$~cm$^{-3}$. The average line-of-sight magnetic
field in the methanol maser region $\langle B_{||, \rm
  meth}\rangle=12$~mG, although this depends on the exact Zeeman
coefficient used to calculate the field strength as discussed in
\S~\ref{circ}. A statistical correction for a randomly oriented
magnetic field gives $|B_{\rm meth}|=23$~mG. This indicates that the
magnetic field is dynamically important. The 100-m Effelsberg
telescope observations presented here, detected significant magnetic
fields in 70\% of the sources with peak fluxes down to
$\sim50$~Jy. Thus, with additional observations it will be possible to
construct a catalogue of magnetic field measurements for over 100
high-mass star-forming regions giving unique insight in Galactic
magnetic fields.

{\it acknowledgments:} WV thanks A.Kraus for his help setting up the
Effelsberg observations, S.Goedhart for making available her most
recent G09.62+0.20 monitoring results and V.Migenes and V.Slysh for
providing details on the IRAS 18089-1732 OH maser observations. WV
also thanks the referee for comments that have greatly
improved the paper.

%\bibliographystyle{aa}
%
%\bibliography{wvrefs}
 \Online

\end{document}